# Validity of the Web-based, Self-directed, NeuroCognitive Performance Test in MCI

P. Murali Doraiswamy[1,2], Terry E. Goldberg[3], Min Qian[4], Alexandra R. Linares[1],

Adaora Nwosu[1], Izael Nino[3,4], Jessica D'Antonio[3,4], Julia Phillips[3,4], Charlie Ndouli[3,4], Caroline Hellegers[1],

Andrew M. Michael[2], Jeffrey R. Petrella[6], Howard Andrews[5], Joel Sneed[7,8], Davangere P. Devanand[3,4]

[1] Neurocognitive Disorders Program, Department of Psychiatry, Duke University School of Medicine, USA

[2] Duke Institute for Brain Sciences, Durham, North Carolina, USA.

[3] Division of Geriatric Psychiatry, New York State Psychiatric Institute, New York, New York, USA.

[4] Department of Psychiatry, Columbia University Medical Center, New York, New York, USA.

[5] Department of Biostatistics, Mailman School of Public Health, Columbia University Medical Center, New York, New York, USA.

[6] Department of Radiology, Duke University Medical Center, Durham, North Carolina, USA.

[7] Department of Psychology, Queens College, City University of New York, Flushing, New York, USA.

[8] Department of Psychology, The Graduate Center, City University of New York, New York, New York, USA.

Correspondence to: adaora.nwosu@duke.edu

Supported by NIA1R01AG052440-01A1. Disclosures are listed at the end.

ABSTRACT

**Background**: Digital cognitive tests offer several potential advantages over established paper-pencil tests but have not yet been fully evaluated for the clinical evaluation of mild cognitive impairment.

**Objective**: The NeuroCognitive Performance Test (NCPT) is a web-based, self-directed, modular battery intended for repeated assessments of multiple cognitive domains. Our objective was to examine its relationship with the ADAS-Cog and MMSE as well as with established paper-pencil tests of cognition and daily functioning in MCI.

**Methods:** We used Spearman correlations, regressions and principal components analysis followed by a factor analysis (varimax rotated) to examine our objectives.

**Results:** In MCI subjects, the NCPT composite is significantly correlated with both a composite measure of established tests (r=0.78, p<0.0001) as well as with the ADAS-Cog (r=0.55, p<0.0001). Both NCPT and paper-pencil test batteries had a similar factor structure that included a large "g" component with a high eigenvalue. The correlation for the analogous tests (e.g. Trails A and B, learning memory tests) were significant (p<0.0001). Further, both the NCPT and established tests significantly (p< 0.01) predicted the University of California San Diego Performance-Based Skills Assessment and Functional Activities Questionnaire, measures of daily functioning.

**Conclusions:** The NCPT, a web-based, self-directed, computerized test, shows high concurrent validity with established tests and hence offers promise for use as a research or clinical tool in MCI. Despite limitations such as a relatively small sample, absence of control group and cross-sectional nature, these findings are consistent with the growing literature on the promise of self-directed, web-based cognitive assessments for MCI.

# INTRODUCTION

The advent of cloud-based online and smartphone cognitive tests has enhanced access to home-based and point of care memory evaluation [1-3]. There has been a recent proliferation of such tests including several that have been cleared by regulatory agencies [4-13]. Such tests offer many potential advantages over conventional clinician-administered neuropsychological tests such as greater consistency of administration, easy generation of alternate forms, 24/7 access on demand, better stimulus control, greater ability to test at home, greater ability to personalize and minimize floor or ceiling effects, automation of scoring, cloud storage and integration with electronic health records, and ability to scale to diverse populations and settings, such as areas with clinician shortages [12]. Indeed, the computerized version of the ADAS-Cog has been shown to have better reliability than the traditional paper-pencil version [13]. Further, the pandemic has also raised the need for remote contactless assessment both for clinical purposes and clinical trials in elderly at-risk patients [1]. There are also potential disadvantages of computerized tests in that they can be difficult for more severely impaired individuals or those unfamiliar with computers and are prone to technical glitches and privacy breaches. These issues argue for further study in diverse clinical settings to determine the best tests for specific purposes [14].

The NeuroCognitive Performance Test (NCPT) is an online, digital platform intended for repeated assessments of multiple domains such as different types of attention, memory, executive functioning, and psychomotor speed [12]. It currently has 18 modules (subtests), each based on well-known neuropsychological assessments and its modular design allows researchers to develop customized batteries to address specific research outcomes. The NCPT is a self-directed test designed to be taken through a web browser. Morrison et al [12] reported on normative data and factor structure of an 8-item NCPT in 130,140 healthy volunteers, drawn from 187 countries and across a wide educational and age range. They reported adequate test-retest reliability over 70 days (in a subset of 35,779 users) as well as good concurrent validity to standard neuropsychological tests (in a subset of 73 younger subjects). Compared to age-, gender-, and education-matched normal controls, the NCPT composite was 0.78 SD lower in subjects with self-reported mild cognitive impairment (MCI ) (N=1,473) and 1.17 SD lower in subjects with self-reported Alzheimer's disease (AD) (N=105) [12]. In a subsequent study of 4715 subjects, a 7-item version of the self-directed NCPT was sensitive to detecting the effects of cognitive training [15]. These findings support further study of the utility of the NCPT in clinical populations.

In this study, we report our cross-sectional experience using the NCPT using baseline data from an ongoing two-site, prospective clinical trial of cognitive brain training MCI - further details of the COG-IT study design have been reported elsewhere [16]. A 10-item version of the NCPT was created for this study incorporating modules thought to be sensitive to the study interventions [16]. The aims of this paper were to evaluate the feasibility of NCPT-10 self-administration and examine its construct against standard paper-pencil tests.

## METHODS

### Subjects and Study design

All subjects gave written informed consent and the study was approved by the respective institutional IRBs. Details of study design, inclusion exclusion criteria and study procedures have been previously reported [16]. 109 MCI subjects were recruited for a prospective clinical trial of computerized brain training at two study sites (New York and Durham) stratified by MCI severity (early MCI or late MCI) and age (70 and below or 71 and above). Only baseline data was used for this analyses. One hundred and one subjects with completed baseline NCPT and standard neuropsychological test data were included in this analysis. A clinical diagnosis of MCI was made after neuropsychiatric and neuropsychological evaluation as described previously [16]. All subjects also underwent brain MRI. Severity of MCI was assessed by the delayed recall score of WMS-III Logical Memory as described previously [16]. Notable inclusion criteria included an age range of 55–95 years, subjective cognitive complaints (ie, memory or other cognitive complaints, eg, naming/language), a Wechsler Memory Scale-III (WMS-III) Logical Memory delayed recall score 0-11 (education adjusted), Mini Mental State Examination (MMSE) score ≥23 out of 30, availability of an informant and access to a home desktop or laptop computer with full access to the Internet for the study duration. Notable exclusion criteria included major neuropsychiatric illness, lack of English-speaking ability, and regular use of brain training games [16].

**Neuropsychological and Functional Assessments**

At baseline, the Alzheimer's Disease Assessment Scale-Cognition Subscale 11 (ADAS-Cog 11) was administered, followed by the following neuropsychological test battery: Digit Symbol Substitution Test (DSST) (to assess attention), WMS-III Visual Reproduction Test (to assess nonverbal learning and memory), Auditory-Verbal Learning Test (to assess verbal learning and memory), Block design, Verbal and Category Fluency, Trail Making A & B (to assess attention and executive function), and 15-item Boston Naming Test (to assess language) [16]. Following this test battery, the University of California San Diego Performance-Based Skills Assessment (UPSA-3) was administered [16-17]. The UPSA-3 is a performance-based measure of cognitive and functional abilities that includes measures of simulated real-world activities; for example, planning a trip to the beach, remembering documents to bring to a medical appointment, and dialing a phone number [17]. The Functional Activities Questionnaire (FAQ) was administered to the patient's informant, either during the study visit or shortly after the visit over the phone. Testing fatigue was mitigated by allowing participants to take breaks during the testing.

**NeuroCognitive Performance Test (NCPT-10)**

Following the paper-pencil testing session, subjects were then taken to a quiet clinic room so they could do the NCPT (Lumos Labs) by themselves in a self-directed manner. The study provided the on-site computer for this test. A research associate was available to help troubleshoot if computer glitches arose. The subtests and cognitive domains measured by the NCPT are described in Supplemental Table 1. In summary they were memory (visuospatial working memory, short-term memory), processing speed (visual search, psychomotor speed), problem solving (logical reasoning, numerical calculation), attention (selective, divided) and flexibility (response inhibition, task switching). The 10 NCPT subtests were online adaptations of widely used neuropsychological tests [12, 16, Supplemental Table 1]. The NCPT took about 20-40 minutes.

**Statistical Approaches**

In this paper our goal was to examine the cross-sectional relationship, at baseline, of established tests (9 domain-specific neuropsychological tests and 2 global cognitive tests [ADAS, MMSE]) to the 10-item NCPT as well as the NCPT's relationship to measures of daily function (the UPSA-3 and FAQ) and disease severity. Summary statistics for demographic and various tests were calculated. Spearman correlation were constructed to examine the interrelations of the various tests. For the NCPT and the paper-pencil tests separately, we conducted a principal components analysis followed by a factor analysis (varimax rotated). For the principal components we determined the number of factors by setting eigen values>1 and by scree plots. We considered an individual test to be associated with a specific factor if its loading was greater than 0.40. Linear and logistic regressions examined the association of the factors with functional measures and disease severity. We also constructed an exploratory composite Z score for the NCPT tests weighting the 10-tests equally, as well as composite z-score for the key paper-pencil tests, and examined their association with the functional measures and disease severity. For computing the composite z-scores for both established and NCPT batteries, we reverse coded (reversing sign) ADAS-Cog, Trails-A and Trails-B, so that higher numbers for all tests reflect better performance. For the NCPT, in most cases, the 3-factor model performs slightly better than or similar to the single composite z-score model in terms of R-square or adjusted R-square. As such, this was an exploratory analysis and we did not adjust the p-values for multiple comparisons.

## RESULTS

Table 1 depicts the baseline characteristics of the 101 MCI subjects. Of these, 44 subjects were classified as early MCI (EMCI) and 57 as late MCI (LMCI). Subjects recruited at Durham were on average older than those recruited in NYC but there were no differences in gender ratio or educational level.

**Correlations between NCPT and Paper-Pencil tests**

Within the NCPT and traditional test batteries, there were multiple significant test intercorrelations, suggesting the potential presence of a general cognitive ability factor (or g) factor in each (data not shown). The correlation between the composite of all paper-pencil tests (CompZ) and the NCPT composite (NCPT-Z) was high (r=0.78, p<0.0001). The correlation for the analogous tests, Trails A (r=0.43) and Trails B (r=0.63) between the NCPT and the paper-pencil versions were significant (p<0.001). Likewise, the correlations between the NCPT word list learning memory tests and paper-pencil AVLT were significant (p<0.0001). The correlations between NCPT composite Z score and ADAS-Cog (r=-0.55, p<0.0001) (Figure 1) and MMSE (r=0.53, p<0.0001) are also shown because the latter tests are used widely in clinical research.

**Principal Components and Varimax Analyses of NCPT and Paper-Pencil tests**

Three factors were derived for both the NCPT and paper-pencil measures (Supplemental Table 2). For the NCPT the first component accounted for 30% of the variance and was comprised of multiple tests from different domains (e.g., scale balance reasoning, trails speed, arithmetic). The second component was comprised of word list learning scores (verbal recall) as well as digit symbol coding and Trails. The third was comprised of the two object recognition memory scores. The three components accounted for 69% of the variance. For the paper-pencil tests, the component factor with the largest eigenvalue included tests of memory, general tests (MMSE and ADAS-Cog) and speed, which could represent "g" (Supplemental Table 2). It accounted for 29% of the variance. A second factor was comprised of high loadings from the AVLT delayed list learning and delayed logical memory. The third factor included logical memory immediate and naming, suggesting a verbal recall construct. All three components accounted for 68% of the variance, which was similar to the 69% for the NCPT.

The first factors in each battery were highly inter-correlated even after adjusting for age, gender and education (r=0.60, p<0.0001). This suggests they both assay a latent construct involving general cognitive ability ("g"). The second factors in each battery were also significantly inter-correlated (r=0.35, p=0.0004) after adjustment for age, education and gender. The third factor in each battery was not inter-correlated (p=0.618).

**Relationship of Factors to Daily Functioning**

We examined the relationship of these factor scores to daily functioning (FAQ, UPSA) and severity of MCI (EMCI vs LMCI). The results below report values for regressions after adjusting for age, sex, and education (Supplemental Table 3). For the NCPT, the UPSA was predicted by factors 1 (g) and 2 ($p < 0.0001$) and 3 (p=0.001), and the FAQ by factors 1 (g) (p=0.002) and 2 (p=0.001). In the factor analyses, none of the three factors for the NCPT predicted early versus late MCI membership since this version of the NCPT was not as weighted to verbal memory tests. (The primary memory test in the NCPT, word list learning total, did significantly separate EMCI from LMCI [p=0.009]). For the established measures, the UPSA was predicted by factors 1, 2, and 3 (p<0.0001) and the FAQ by factors 1 and 2 (p<0.0001). For the paper-pencil measures, which comprised multiple verbal memory tests, MCI severity was predicted by factors 2 and 3 (p<0.0001). For the two functional measures (UPSA and FAQ), the $R^2$ accounted for by both NCPT and paper-pencil tests were much larger than that accounted for by demographics. Alternate models with all factors or 5-factors were also examined but not deemed superior (data not shown). A separate regression analyses (not using principal factors) confirmed that NCPT and paper-pencil composite scores both significantly predicted the UPSA and FAQ (data not shown).

**DISCUSSION**

The NCPT, a self-directed, web-based, computerized test, has one of the largest normative databases of its kind derived from over 40 million users in a natural setting [12, 18]. However, in contrast to some other computerized tests, such as NTB [19], CANTAB [9], Cogstate [6] or CNS Vital Signs [7], it has not been used as extensively in memory clinics or in MCI/AD clinical trials. In this analysis, we compared the NCPT with a paper and pencil battery in subjects with clinically diagnosed MCI.

Several key findings emerged from our study. The feasibility of completing the self-directed NCPT was high among MCI subjects. Both test batteries had a similar factor structure that included a large "g" component with a high eigenvalue. Both test batteries also showed broadly similar factor score prediction of two clinically meaningful functional outcomes. Both NCPT and established test batteries were also significantly associated with FAQ and UPSA, which are established, clinically meaningful, functional measures [17]. Analogous tests in the NCPT and paper-pencil batteries, such as Trails and word list learning, were highly correlated. Likewise, all ten NCPT tests as well as an NCPT z-score composite were significantly correlated with the ADAS-Cog, an instrument used widely in clinical trials. All ten NCPT items and the composite were also correlated with the MMSE, a screening tool widely used in clinical practice. These data support the concurrent validity of the NCPT.

The strengths of our study are the clinical verification of MCI diagnosis and standardized administration of a comprehensive battery of neuropsychological tests on the same day as the NCPT. However, there were also some limitations to our study. Our analysis was cross-sectional and lacked a healthy control group; hence, we could not directly assess the utility of NCPT as a diagnostic or prognostic tool. The NCPT modules in this study were selected to examine intervention effects and as such were not exactly analogous to the paper-pencil tests. This was taken into account in conducting the factor analyses and did not affect the results materially with factor 1 and factor 2 being similar for the paper-pencil battery and NCPT. We did not measure biomarkers and hence cannot determine the cause of MCI. While the feasibility was high among MCI subjects, our study included a trouble-shooting monitor – hence, further study is needed to confirm if the feasibility in unsupervised home settings is equally high.

In summary, our analysis finds that the NCPT, a web-based, self-directed, computerized test, shows high concurrent validity with established tests and hence offers promise for use as a research or clinical tool in MCI. Given the increasing numbers of smart phone, wearable, voice, and web-based computerized tests now available to assess cognition [1-4], it will also be important to directly compare these newer tests against each other at various disease stages. Such studies will allow clinicians to better personalize the tests needed for specific clinical assessments and research questions.

**ACKNOWLEDGEMENTS**

**Funding:** This work is supported by National Institute on Aging grant (1R01AG052440-01A1). We thank Lumos Labs for providing the gaming platform at no cost but they had no involvement in the analyses or decision to publish the study.

**Competing interests:** PMD has received grants from NIA, DARPA, DOD, ONR, Salix, Avanir, Avid, Cure Alzheimer's Fund, Karen L. Wrenn Trust, Steve Aoki Foundation, and advisory fees from Apollo, Brain Forum, Clearview, Lumos, Neuroglee, Otsuka, Verily, Vitakey, Sermo, Lilly, Nutricia, and Transposon. PMD is a co-inventor on patents for diagnosis or treatment of Alzheimer disease as well as a patent for infection detection. PMD owns shares in several biotechnology companies whose products are not discussed here. DPD serves as a consultant on advisory boards to Acadia, Biogen, BioExcel, Genentech, Eisai, GW Pharmaceuticals, and Novo Nordisk. Other authors have received grant support from NIH and/or industry studies but report no other competing interests.

**Table 1. Baseline Characteristics of MCI Subjects***

| Label | Total |
|---|---|
| N | 101 |
| NYC / Durham | 52 / 49 |
| Age (y) | 70.8 $\pm$ 8.9 |
| Female/Male (%) | 58.4 / 41.6 |
| White / Minority (%) | 76.2 / 23.8 |
| Education (y) | 16.7 $\pm$ 3.1 |
| EMCI / LMCI | 44 / 57 |
| MMSE Total | 27.1 $\pm$ 1.7 |
| Wechsler Logical Memory Delay | 6.4 $\pm$ 3.3 |
| ADAS-Cog-11 Total | 9.5 $\pm$ 3.5 |
| UPSA Total | 80.9 $\pm$ 11.3 |
| FAQ Total | 3.4 $\pm$ 4.0 |

(mean $\pm$ SD). *Please see text for details on sample size and abbreviations.

**Figure 1: Correlation between NCPTz and ADAS-Cog in MCI**

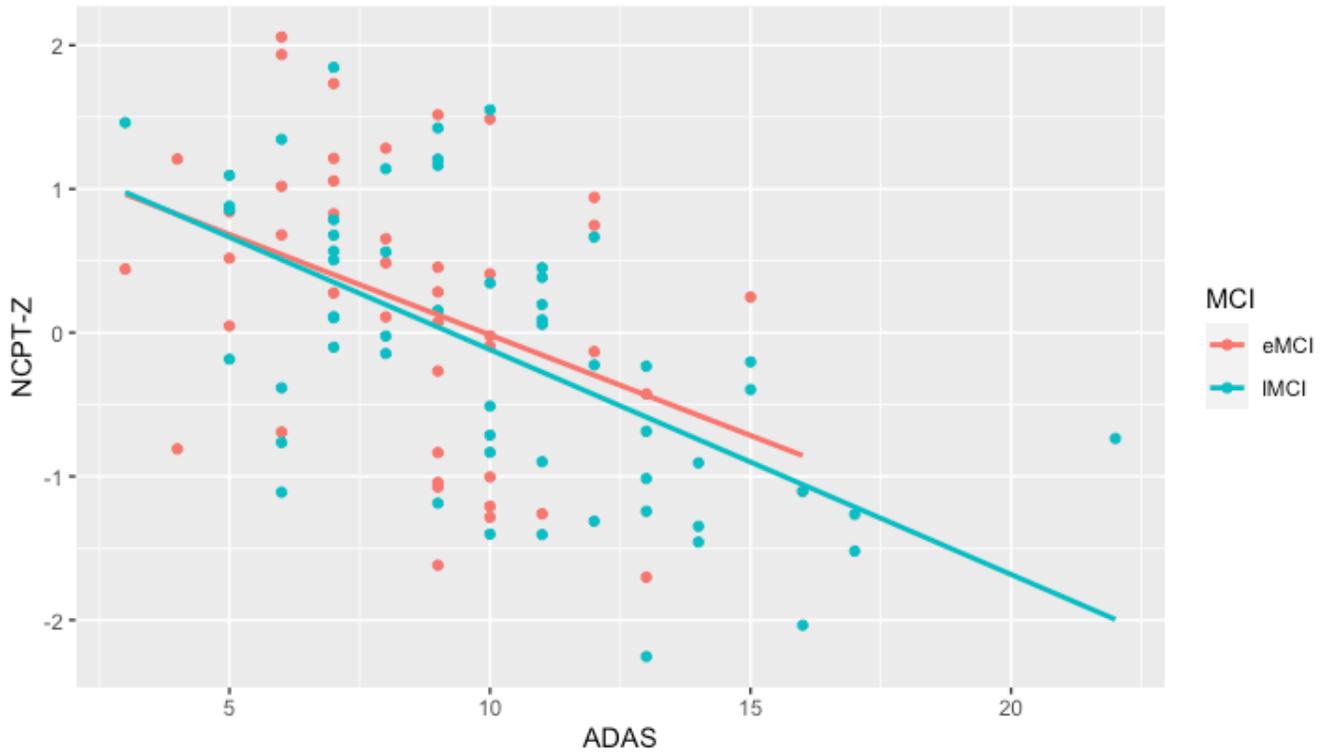

Figure 1 shows correlation between NCPT-Z composite and ADAS-Cog. The red and blue regression lines show correlations for EMCI and LMCI subjects respectively. Raw ADAS-Cog scores (not reverse coded) were used for this figure. See text for details.

**Supplemental Table 1. Web-Based, Self-Directed, NCPT Modules Used in This Study**

- **Word List Learning.** In this assessment, subjects are presented a list of words and then type the words from free recall. The dependent measure is the total number of correct responses. Measure immediate recall.

- **Digit Symbol Coding.** Digit Symbol Coding, based on the Digit Symbol Substitution Task, is a measure of information processing speed. The participant uses the number keyboard to type the number that corresponds to the symbol. The dependent measure is the total number of correct trials minus incorrect trials over 90 seconds. Measures processing speed and memory.

- **Scale Balance.** Subjects are required to determine which solution, represented by a set of shapes, balances a scale. The dependent measure is the total number of correct responses. Measures problem solving.

- **Word List Learning Delayed.** In this assessment, subjects recall as many words as possible from the initial Word List. The dependent measure is the total number of correct responses. Measures delayed recall.

- **Object Recognition.** Subjects remember a series of images and make a forced-choice based on whether an image was previously presented. The dependent measure is the total number of correct responses. Test of visual memory.

- **Arithmetic Reasoning.** A cognitive task in which individuals solve basic arithmetic questions, written in words. The total correct responses over 90 seconds is the dependent measure. Measures problem solving ability.

- **Memory Span (Forward & Reverse).** A common measure of visual short-term memory derived from the Corsi block-tapping test. The session ends when the participant gives three incorrect answers at the same span level. The total number of correct responses is the dependent measure.

- **Trail Making A.** In Trail Making A, subjects are required to click on the numbered circles in sequential order as quickly as possible. The dependent variable is the completion time for the task. Used to measure speed of processing.

- **Trail Making B.** Trail Making B tests additional cognitive domain of mental flexibility. The dependent variable is the completion time for the task. Used to measure processing speed and mental flexibility.

- **Object Recognition Delayed.** Subjects are presented with two objects at a time and required to select which image appeared in the original list. The dependent measure is the total correct responses. Test of delayed visual memory.

**Supplemental Table 2. Factor Loadings of PCA with Varimax Rotation for established and NCPT measures**

|  | Factor1 | Factor2 | Factor3 |
|---|---|---|---|
| **Paper-Pencil tests** |  |  |  |
| ADAS-Cog | 0.56 | 0.43 | 0.19 |
| MMSE | 0.62 | 0.44 | 0.01 |
| Logical Memory immediate | 0.04 | 0.47 | 0.71 |
| Logical Memory delay | 0.22 | 0.6 | 0.55 |
| WAIS-R DSST (DSST) | 0.81 | 0.22 | 0.06 |
| AVLT Short Delay (AVLT-I) | 0.22 | 0.87 | 0.08 |
| AVLT Long Delay (AVLT-II) | 0.24 | 0.89 | 0.02 |
| Category Fluency | 0.57 | 0.16 | 0.38 |
| Boston Naming Test | 0.2 | -0.14 | 0.79 |
| Trail Making A | 0.8 | 0.11 | 0.16 |
| Trail Making B | 0.84 | 0.09 | 0.12 |
|  |  |  |  |
| **NCPT Tests** |  |  |  |
| Object Recognition Delay | 0.23 | 0.19 | 0.83 |
| Digit Symbol Coding | 0.61 | 0.54 | 0.27 |
| Scale Balance | 0.65 | 0.19 | 0.23 |
| Object Recognition Immediate | 0.29 | 0.21 | 0.79 |
| Memory Span | 0.8 | 0.03 | 0.06 |
| Trail Making A | 0.46 | 0.58 | -0.31 |
| Trail Making B | 0.72 | 0.2 | 0.23 |
| Word List Learning | 0.29 | 0.8 | 0.26 |
| Word List Learning Delay | -0.03 | 0.83 | 0.33 |
| Arithmetic Reasoning | 0.76 | 0.16 | 0.23 |

Table depicts factor loadings from principal component analyses with varimax rotation of 3 factors with eigen values >1. Shading indicates that factor loading was greater than 0.40. Please see text for details.

**Supplemental Table 3a: Linear/logistic regression models based on 3 rotated factor models for NCPT measures.**

| | UPSA (coef and p-value) | | | FAQ (coef and p-value) | | | eMCI (odds ratio and p-value) | | |
|---|---|---|---|---|---|---|---|---|---|
| Variable | Demo only | Factor only | full | Demo only | Factor only | full | Demo only | Factor only | full |
| Intercept | 107.32 (p= 0.000) | 80.90 (p= 0.000) | 79.95 (p= 0.000) | -1.41 (p= 0.739) | 3.41 (p= 0.000) | 6.62 (p= 0.132) | 2.76 (p= 0.647) | 0.76 (p= 0.190) | 0.47 (p= 0.763) |
| Factor1 | | 4.76 (p= 0.000) | 4.54 (p= 0.000) | | -1.31 (p= 0.000) | -1.29 (p= 0.002) | | 1.14 (p= 0.534) | 1.20 (p= 0.437) |
| Factor2 | | 4.88 (p= 0.000) | 4.63 (p= 0.000) | | -1.37 (p= 0.000) | -1.41 (p= 0.001) | | 1.41 (p= 0.103) | 1.40 (p= 0.157) |
| Factor3 | | 3.09 (p= 0.000) | 2.95 (p= 0.001) | | -0.49 (p= 0.173) | -0.46 (p= 0.210) | | 1.18 (p= 0.443) | 1.18 (p= 0.438) |
| sex (ref. male) | 1.07 (p= 0.626) | | 1.95 (p= 0.286) | -0.63 (p= 0.442) | | -0.92 (p= 0.222) | 1.35 (p= 0.485) | | 1.42 (p= 0.421) |
| Age | -0.50 (p= 0.000) | | -0.07 (p= 0.583) | 0.12 (p= 0.009) | | -0.01 (p= 0.899) | 0.98 (p= 0.381) | | 1.01 (p= 0.860) |
| edu | 0.46 (p= 0.179) | | 0.15 (p= 0.614) | -0.15 (p= 0.226) | | -0.08 (p= 0.530) | 0.98 (p= 0.786) | | 0.97 (p= 0.724) |
| R-Square | 0.171 | 0.44 | 0.451 | 0.087 | 0.239 | 0.252 | Pseudo R-square =0.017 | 0.037 | 0.045 |
| Adj R-Sq | 0.146 | 0.423 | 0.416 | 0.059 | 0.215 | 0.204 | AIC=144.59 | 142.54 | 147.66 |

**Supplemental Table 3b: Linear/logistic regression models based on 3 rotated factor models for paper-pencil measures.**

| | UPSA (coef and p-value) | | | FAQ (coef and p-value) | | | eMCI (odds ratio and p-value) | | |
|---|---|---|---|---|---|---|---|---|---|
| Variable | Demo only | Factor only | full | Demo only | Factor only | full | Demo only | Factor only | full |
| Intercept | 107.32 (p= 0.000) | 80.90 (p= 0.000) | 98.71 (p= 0.000) | -1.41 (p= 0.739) | 3.41 (p= 0.000) | 3.10 (p= 0.394) | 2.76 (p= 0.647) | 0.50 (p= 0.026) | 1694.9 (p= 0.051) |
| Factor1 | | 6.69 (p= 0.000) | 6.19 (p= 0.000) | | -2.05 (p= 0.000) | -2.12 (p= 0.000) | | 1.29 (p= 0.380) | 1.50 (p= 0.272) |
| Factor2 | | 3.71 (p= 0.000) | 3.38 (p= 0.000) | | -1.31 (p= 0.000) | -1.35 (p= 0.000) | | 4.19 (p= 0.000) | 8.99 (p= 0.000) |
| Factor3 | | 4.22 (p= 0.000) | 4.46 (p= 0.000) | | -0.65 (p= 0.041) | -0.65 (p= 0.047) | | 11.11 (p= 0.000) | 31.52 (p= 0.000) |
| sex (ref. male) | 1.07 (p= 0.626) | | -0.37 (p= 0.808) | -0.63 (p= 0.442) | | -0.24 (p= 0.719) | 1.35 (p= 0.485) | | 0.83 (p= 0.773) |
| Age | -0.50 (p= 0.000) | | -0.19 (p= 0.051) | 0.12 (p= 0.009) | | -0.01 (p= 0.831) | 0.98 (p= 0.381) | | 0.98 (p= 0.652) |
| edu | 0.46 (p= 0.179) | | -0.23 (p= 0.342) | -0.15 (p= 0.226) | | 0.08 (p= 0.467) | 0.98 (p= 0.786) | | 0.66 (p= 0.004) |
| R-Square | 0.171 | 0.6 | 0.623 | 0.087 | 0.396 | 0.401 | Pseudo R-square =0.017 | 0.452 | 0.513 |
| Adj R-Sq | 0.146 | 0.587 | 0.599 | 0.059 | 0.377 | 0.363 | AIC=144.59 | 85.55 | 79.73 |